\documentclass[prb,twocolumn,superscriptaddress,showpacs,amsmath,amssymb]{revtex4-2}
\usepackage{graphicx}
\usepackage{amsmath}
\usepackage{amssymb}
\usepackage{float}
\usepackage{color}

\usepackage{cancel,xcolor}
\usepackage{indentfirst}
\usepackage{CJKulem}
\usepackage{soul,xcolor}
\setstcolor{red}

\usepackage{ulem}
\usepackage{cancel}
\usepackage[pdfpagemode=UseNone,pdfstartview=FitH,colorlinks=true,linkcolor=blue,urlcolor=blue,anchorcolor=blue,citecolor=blue]{hyperref}

\newcommand{\beq}{\begin{equation}}
\newcommand{\eeq}{\end{equation}}
\newcommand{\beqa}{\begin{eqnarray}}
\newcommand{\eeqa}{\end{eqnarray}}

\def\Aa2#1{\textcolor{magenta}{#1}}

\def\Aa1#1{\textcolor{blue}{#1}}
\def\prb#1{{ Phys.\ Rev. B\/} {\bf#1}}

\def\prl#1{{ Phys.\ Rev.\ Lett.} {\bf#1}}

\begin{document}

\title{Bi- and tetracritical phase diagrams in three dimensions}

\author{Amnon Aharony}
\email{aaharonyaa@gmail.com}
\affiliation{ School of Physics and Astronomy, Tel Aviv University, Tel Aviv 6997801, Israel}

\author{Ora Entin-Wohlman}
\email{orawohlman@gmail.com}
\affiliation{ School of Physics and Astronomy, Tel Aviv University, Tel Aviv 6997801, Israel}

\author{Andrey Kudlis}
\email{andrewkudlis@gmail.com}
\affiliation{ITMO University, Kronverkskiy prospekt 49, Saint Petersburg 197101, Russia}

\begin{abstract}

The critical behavior of many physical systems involves two competing $n^{}_1-$ and $n^{}_2-$component order-parameters, ${\bf S}^{}_1$ and ${\bf S}^{}_2$, respectively, with $n=n^{}_1+n^{}_2$.  Varying an external  control parameter $g$, %(e.g. uniaxial stress or magnetic field),
one encounters ordering of ${\bf S}^{}_1$ below a critical (second-order) line for $g<0$ and of ${\bf S}^{}_2$ below another critical line for $g>0$.  These two ordered phases are separated by a first-order line, which meets the above critical lines at a bicritical point, or by an intermediate (mixed) phase, bounded by two critical lines, which meet the above critical lines at a tetracritical point.  For  $n=1+2=3$, the critical behavior around the (bi- or tetra-) multicritical point either belongs to the universality class of a non-rotationally invariant (cubic or biconical) fixed point, or it has a fluctuation driven first-order transition. These asymptotic behaviors arise only very close to the transitions. We present accurate  renormalization-group flow trajectories yielding the  effective crossover exponents near multicriticality.

\end{abstract}

%Remarkably
%%%%%%%%%%%%%%%%%%%%%%%%%%%%%%%%%%%%%%%%%%%%%%%%%
%%%%%%%%%%%%%%%%%%%%%%%%%%%%%%%%%%%%%%%%%%%%%%%%%

\date{\today}


\maketitle
%%%%%%%%%%%%%%%%%%%%%%%%%%%%%%%%%%%%%%%%%%%%%%%%%
%%%%%%%%%%%%%%%%%%%%%%%%%%%%%%%%%%%%%%%%%%%%%%%%%

\section{Introduction}

Many physical systems exhibit critical behavior
which depends on the interplay between two
order-parameters, ${\bf S}^{}_1$ and ${\bf S}^{}_2$, with $n^{}_1$ and $n^{}_2$ components, respectively,  and $n=n^{}_1+n^{}_2$.   A well-studied
example is that of a uniaxially anisotropic antiferromagnet in a uniform magnetic field~\cite{1,2,king,Shapira,NATO},  which
may order antiferromagnetically, with $n^{}_1=1$ (for small values
of the field) or have a spin-flopped phase, with $n^{}_2=2$ (for large values
of the field), with  possibly an intermediate spin-flopped phase. Other examples are associated with
the competition between superfluid and crystal ordering, with a possible intermediate supersolid phase,
in crystalline $^4$He~\cite{4}, between ferroelectric
and ferromagnetic ordering in certain crystals~\cite{5},
between two types of magnetic ordering in mixed
magnetic crystals, e. g., (Mn,Fe)WO$^{}_4$ or Fe(Pd,
Pt)$^{}_3$~\cite{6,AAx,AAF} and between rotations around cubic axes or diagonals
 that characterize the displacive phase transitions
in perovskite crystals~\cite{KAM1970,STO-111,10}.
More recent examples, with larger values of $n$,  concern the competition between an isotropic antiferromagnet ($n^{}_1=3$) and a superconductor ($n^{}_2=2$), relevant for the high-temperature superconductors~\cite{zhang}, between spin-density wave and induced local moments~\cite{mineev}, between superconductivity and spin density waves ($n^{}_1=n^{}_2=2$)~\cite{kivelson} and  in the
temperature baryon-chemical-potential phase diagram of
hadronic matter, with the competing $n^{}_1=4$ order
parameter for chiral symmetry-breaking and the $n^{}_2=6$
order-parameter for color symmetry-breaking~\cite{C8}.

\begin{figure*}[htb]
\vspace{-1.9cm}
\includegraphics[width=1.1\textwidth]{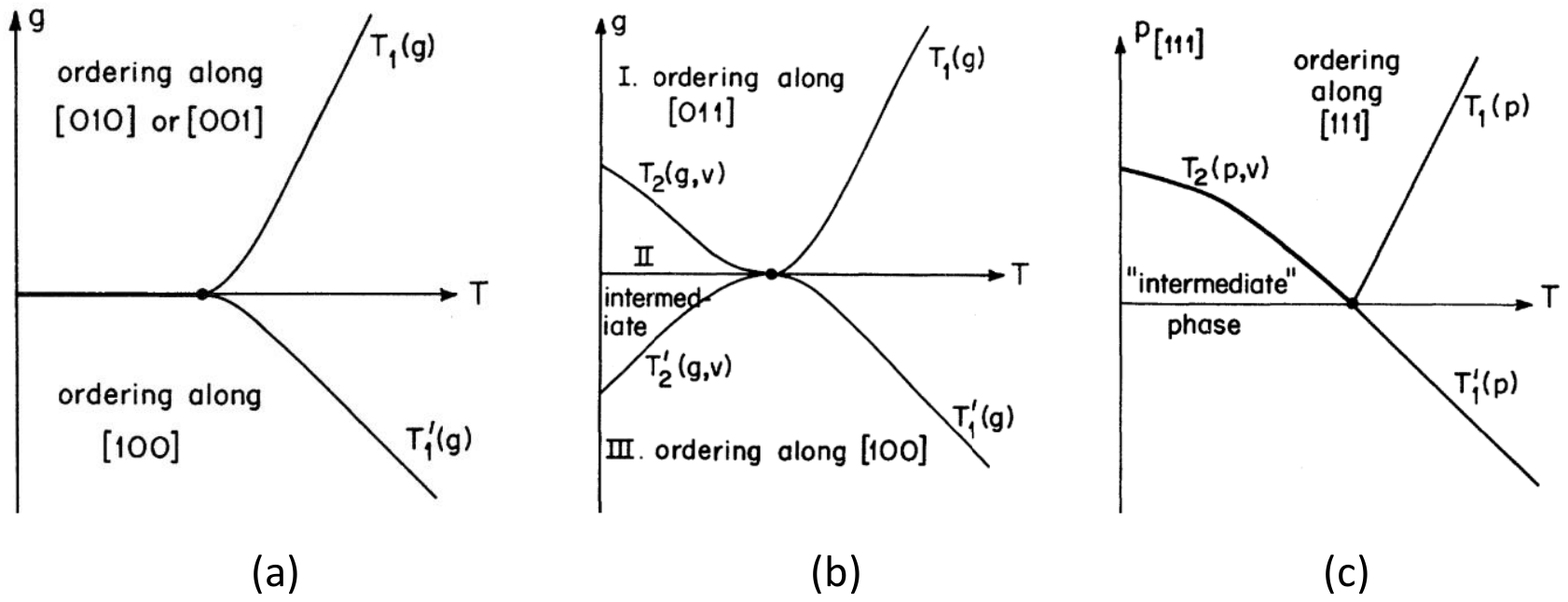}
\vspace{-19.5cm}
\caption{(color online)   Phase diagrams for systems with cubic symmetry. (a) $v<0$ and $g$ appears in Eq. (\ref{Hg}) for $n^{}_1=1,~n^{}_2=2$. Since both $v$ and $g$ prefer ordering along a cubic axis, there is no competition, and the result is  a bicritical phase diagram.  (b) $v>0$. Now $g$ and $v$ compete, prefering ordering along a cubic axis and along a cubic diagonal, and the result is a tetracritical phase diagram. (c) $v<0$ and $p$ appears in Eq. (\ref{Hdiag}).  $p>0$ prefers ordering along $[111]$ but $v<0$ prefers ordering along a cubic axis. The result is a irst-order line corresponding to the thre state Potts model. After Ref. \onlinecite{bruce}.}
\label{1}
\end{figure*}

The nature of the ordered phases in such systems depends on an external tunable parameter, $g$, which couples to the traceless quadratic 
symmetry-breaking Hamiltonian
\begin{align}
{\cal H}^{}_{g}=g\big(|{\bf S}^{}_1|^2-\frac{n^{}_1}{n^{}_2}|{\bf S}^{}_2|^2\big),
\label{Hg}
\end{align}
where ${\bf S}={\bf S}^{}_1+{\bf S}^{}_2$ and
\begin{align}
|{\bf S}^{}_1|^2=\sum_{i=1}^{n^{}_1}(S^{}_i)^2,\ \ \ |{\bf S}^{}_2|^2=\sum_{i=n^{}_1+1}^{n}(S^{}_i)^2.
\label{Hg'}
\end{align}
This Hamiltonian prefers ordering of ${\bf S}^{}_1$ (${\bf S}^{}_2$) for $g<0$ ($>0$).
 For the anisotropic antiferromagnet in a magnetic field, ${\bf S}^{}_1$ is a one-component staggered magnetization, along the easy axis, with $H$ along the same direction, $g$ is a linear combination of $H^2$ and the temperature $T$~\cite{MEFbi}. For the perovskites, $g$ is the uniaxial stress~\cite{10}.
 Figure \ref{1} shows  examples for $n^{}_1=1$ and $n^{}_2=2$.  In Figs. \ref{1}(a,b), ${\bf S}^{}_1$ orders along $[100]$ at $T<T'^{}_1(g),~g<0$, while
 ${\bf S}^{}_2$ orders in the perpendicular plane (along $[010]$ or $[001]$) [panel (a)] or along $[011]$ [panel (b)]) at $T<T^{}_1(g),~g>0$ ~\cite{bruce}.
The two ordered phases are separated by a first-order transition line, at $g=0$ [panel (a)], or by an intermediate (mixed) phase, bounded by two second-order lines, in which both order-parameters are non-zero. The multicritical point at which all the transition lines meet is called a bicritical point [case (a)] or a tetracritical point [case (b)]. In the examples drawn in Fig. \ref{1}(a,b) the details of the ordered phases  are due to cubic symmetry, with the Hamiltonian
\begin{align}
{\cal H}^{}_{cubic}=v\sum_{i=1}^n S_i^4,
\label{Hv}
\end{align}
 which prefers ordering along a cubic axis when $v<0$ [Fig. \ref{1}(a)] or along a cubic diagonal when $v>0$ [Fig. \ref{1}(b)]. In the latter case, ${\cal H}^{}_g$ and ${\cal H}^{}_{cubic}$ prefer competing ordering, hence the intermediate phase.  This competition does not occur in the former case.

 The situation is more complicated when the quadratic anisotropy prefers ordering along a cubic diagonal, e.g. $[111]$. Equation (\ref{Hg}) is then replaced by
 \begin{align}
 {\cal H}^{}_{diag}=-p[S^{}_1S^{}_2+S^{}_1S^{}_3+S^{}_2S^{}_3]/3.
 \label{Hdiag}
 \end{align}
 For $v>0$ and $p>0$, both ${\cal H}^{}_{diag}$ and ${\cal H}^{}_{cubic}$ prefer ordering along a diagonal, e.g., $[111]$. For $p<0$, ${\cal H}^{}_{diag}$ prefers ordering in the plane perpendicular to $[111]$. As soon as there is ordering in that plane,  ${\cal H}^{}_{cubic}$ also generates non-zero values of the $[111]$ component, and therefore both order-parameters order for $p<0$. This mixed phase meets the $[111]-$ordered phase at $g=0$, via a first-order transition line, similar to Fig. \ref{1}(a) (but with $g=-p$ and with the intermediate phase replacing the ordering along $[010]$ and/or $[001]$). For $v<0$ the mixed phase extends to positive $p$, and the ordering at the first-order transition line is related to the three-state Potts model, for the three degenerate directions (at angles $120^o$ with each other) in the plane perpendicular to $[111]$, as shown in
  Fig. \ref{1}(c)~\cite{STO-111,mukDF}.

In the absence of the cubic symmetry, the quartic terms associated with the competition between ${\bf S}^{}_1$ and ${\bf S}^{}_2$ in the Hamiltonian are
\begin{align}
{\cal H}^{}_{n^{}_1-n^{}_2}=u^{}_1|{\bf S}^{}_1|^4+u^{}_2|{\bf S}^{}_2|^4+2w|{\bf S}^{}_1|^2|{\bf S}^{}_2|^2.
\label{n1n2}
\end{align}
The resulting phase diagrams are similar to those in Fig. \ref{1}(a,b), except that now the transitions at $T^{}_1(g)$ and $T'^{}_1(g)$ are into rotationally-invariant phases, with $O(n^{}_2)$ and $O(n^{}_1)$ symmetries (rotational invariance in $n^{}_1-$ and $n^{}_2$-dimensional spaces), respectively.
The bicritical (tetracritical) phase diagram arises for $w^2>u^{}_1u^{}_2$ ($w^2<u^{}_1u^{}_2$)~\cite{13}.

%\begin{widetext}

%\end{widetext}

Early renormlization-group studies of such multicritical phase diagrams appeared in the early 1970'ies~\cite{13,14,bruce,DG}.
The critical behaviors along the lines $T^{}_1(g)$ and $T'^{}_1(g)$ belong to the universality classes of the $n^{}_2$-component and the $n^{}_1$-component stable fixed points, respectively. For $n^{}_2=2$ and $n^{}_1=1$ in $d=3$ dimensions these are the isotropic XY and Ising model critical behaviors. However, the critical behavior near the $n$-component multicritical point has been under dispute for many years.
Within the $\epsilon-$expansion, in $d=4-\epsilon$ dimensions, the renormalization-group studies found that the $n-$component critical behavior is described by the isotropic (Heisenberg-like) fixed point only for $n<n^{}_c(d)$, where $n^{}_c(d)$ is a borderline number of order-parameter components. Early calculations, to a low order in $\epsilon$, yielded $n^{}_c(3)>3$, and therefore the isotropic behavior prevailed at $d=3$~\cite{NATO}.  Consequences from this assumption appear e.g. in Refs. \onlinecite{bruce,13,14,DG}.  However, at $d=3$ two other fixed points, i.e. the cubic fixed point (for ${\cal H}^{}_{cubic}$) \cite{AA1973,DG} and the biconical fixed point~\cite{fn} (for ${\cal H}^{}_{n^{}_1-n^{}_2}$) \cite{14}, were not very far from this fixed point, and became more stable as $n$ became larger than $n^{}_c(d)$.

Later accurate calculations, both for the cubic case~\cite{MC,6loops,eps6,boot,vic-rev} and for the Hamiltonian (\ref{n1n2}) case~\cite{MC,vicari,folk,vic-rev}, found that $2.85<n^{}_c(3)\lessapprox 3$. In the cubic case, the cubic fixed point is stable for all $n\geq 3$. In the (\ref{n1n2}) case the biconical fixed point is stable for $n^{}_1+n^{}_2=3$, but the decoupled fixed point, in which asymptotically the two order-parameters maintain the critical behaviors which they had for $g\ne 0$, becomes stable for
$n\geq 4$. In fact, this statement follows from  exact scaling arguments~\cite{AAx,DG,AA,AAF}, which do not depend on the $\epsilon-$expansion. Although in some cases this decoupling seems to occur only very close to the multicritical point~\cite{AA}, these scaling arguments resolve the problem for $n^{}_1+n^{}_2=n>3$. The two critical lines simply cross each other, and the phase diagram is tetracritical. The situation for $n=2$ is also simple, and the two Ising-like critical lines meet at the isotropic XY model multicritical point. The cubic interaction is then dangerously irrelevant, since its sign still determines the details of the multicritical phase diagram. However, the most common case $n=3=1+2$ still requires more discussion.

Since most of the earlier theoretical discussions of the multicritical point at $n=d=3$ were based on the assumption that this point has the critical behavior of the isotropic Heisenberg fixed point, the present paper aims to discuss the theoretical changes required by the fact that this fixed point is replaced by the cubic or the biconical fixed points.
 Recently~\cite{AEK} we have discussed the consequences of this instability for the cubic case, for $g=0$.  That paper contains a detailed review of the renormalization-group analysis in the cubic case. However, we have not discussed the detailed consequences on the multicritical phase diagrams, like those shown in Fig. \ref{1}. The present paper performs the latter task for the cubic case, and also includes some discussion of multicritical phase diagrams with $n=1+2$ order-parameter components.

If one starts very close to the phase transitions then we conclude that one should either observe the tetracritical phase diagram associated with the stable cubic or biconical fixed points, or a generalized bicritical phase diagram, in which the bicritical point is replaced by a triple point at which three first-order lines meet. However, the slow renormalization-group flow near the (unstable)  isotropic fixed point implies that under most conditions one will only find effective phase diagrams, which are similar to the bicritical diagram, Fig. \ref{1}(a), with effective exponents. These predictions seem to be confirmed by many experiments.

Our renormalization-group calculations  of the critical exponents associated with the anisotropies  (\ref{Hg}) and (\ref{Hdiag}), in the cubic case,  are summarized in
Sec. II. The implications of these results for the multicritical phase diagrams in this case are then discussed in Sec. III. Section IV contains a general discussion of quartic symmetry-breaking terms, and the implications of such terms for the multicritical phase diagrams in the case (\ref{n1n2}), with the stable biconical fixed point. Comparisons with experiments are listed in Sec. V, and our conclusions are summarized in Sec. VI.

%we present the most general Hamiltonian which describes such phase diagrams, and describe the renormalization-group analysis of this Hamiltonian. Since all the current literature agrees that the isotopic, cubic and biconical fixed points are all very close to each other, we show that the approximation in which all three fixed points are identical at $n=d=3$ is quite good. Given that, Sec. III describes the renormalization-group flows in the vicinity of this common fixed point, with emphasis on the parts relevant to Fig. \ref{1}. As discussed in Ref. \onlinecite{AEK}, the transition at the multicritical point may be second order or fluctuation driven first-order. In both cases, the renormalization-group flow is slow, and therefore the critical behavior exhibits effective exponents, which are also important in determining the shapes of all the transition lines in Fig. \ref{1}. This is particularly important for the fluctuation driven first-order case.

\section{The cubic case}

As reviewed in Ref.~\onlinecite{AEK}, the starting point of the calculation is the isotropic Wilson-Ginzburg-Landau normalized free-energy (in dimensionless units),
 $\int d^dr {\cal H}({\bf r})$,  where
\begin{align}
{\cal H}({\bf r})\equiv|{\boldmath{\nabla}}{\bf S}({\bf r})|^2/2+r|{\bf S}({\bf r})|^2/2+u |{\bf S}({\bf r})|^2.
\label{H0}
\end{align}
For the two anisotropic cases this Hamiltonian is augmented by Eqs. (\ref{Hv}) or (\ref{n1n2}).
To discuss the multicritical phase diagrams,  the quadratic anisotropies (\ref{Hg}) or (\ref{Hdiag}) are added as well.

Within the renormalization-group scheme~\cite{wilson,WK,MEF1,MEF,WF,DG}   one first eliminates the short-length details, on scales below $1/e^\ell$ ($\ell$  counts the number of iterations in the elimination process and the lengths are normalized by the unit cell size).
Rescaling the unit length by the factor $e^\ell$ yields a renormalized effective (dimensionless) Hamiltonian (or free-energy density) ${\cal H}(\ell)$, which `flows' in the space spanned by all such Hamiltonians. These flows represent the renormalization-group.
 Critical points are associated with fixed points of these flows, which are invariant under the renormalization-group iterations.  Near a fixed point,
a perturbation $\mu^{}_i(0){\cal O}^{}_i[{\bf S}]$, where ${\cal O}^{}_i[{\bf S}]$ is some polynomial in the components of ${\bf S}$ and $\mu^{}_i$ is a scaling field,  is renormalized as $\mu^{}_i(0)\rightarrow \mu^{}_i(\ell)=\mu^{}_i(0)e^{\lambda^{}_i\ell}$.
A stable fixed point has only two relevant scaling fields, the temperature $\mu^{}_1=t=T/T^{}_c-1$ (which is related to $r$ after a shift due to fluctuations)
and $\mu^{}_2={\bf h}$ (related to the ordering field, ${\bf h}\cdot{\bf S}({\bf r})$). These have positive exponents, $\lambda^{}_1=1/\nu$ and $\lambda^{}_2=(\beta+\gamma)/\nu$, where $\nu,~\beta$ and $\gamma$ correspond to the singular behavior  of the correlation length, $\xi\sim |t|^{-\nu}$,  the order-parameter, $|\langle {\bf S}\rangle|\sim |t|^\beta$  and the susceptibility, $\chi\sim |t|^{-\gamma}$. Other critical exponents are found by scaling relations. Below we discuss only the case $h=0$, and include only even polynomials in ${\bf S}$. All other perturbations are irrelevant, with $\lambda^{}_i<0,~i\geq 3$. An unstable fixed point has more relevant scaling fields, with positive `stability exponents' $\lambda^{}_i>0$.

The renormalization-group for the cubic Hamiltonian~\cite{AA1973,DG,BC1973,Wallace1973} yielded four fixed points in the $u-v$ plane:, the Gaussian, $u^\ast_G=v^\ast_G=0$, the  isotropic, $v^{\ast}_I=0,~u^\ast_I>0$, the decoupled Ising, $u^{\ast}_D=0,~v^\ast_D>0$ (for which the different $S^{}_i$'s decouple from each other and exhibit  the Ising model behavior),  and  the `cubic' fixed points. The location of the latter, $(u^{\ast}_C,~v^{\ast}_C)$,  depends on the number of the order-parameter components,
 $n$: for small (large) $n$, it is in the lower (upper) half $u-v$ plane. It is now accepted that the cubic fixed point is stable, and that all initial Hamiltonians with $u(0),~v(0)>0$ flow to it on the critical $u-v$ plane (on which $t=0$ and the correlation length is infinite).

To find accurate renormalization-group recursion relations in $d=3$, i.e., $\epsilon=1$, we have used order 6 $\epsilon-$expansions at the isotropic fixed point~\cite{eps6}  to obtain  expansions of these recursion relations to quadratic order in $v(\ell)$ and $\delta u(\ell)=u(\ell)-u^\ast_I$~\cite{AEK}. The coefficients in these quadratic expansions were derived using a resummation of the (divergent) $\epsilon-$expansions at $n=3$ and $\epsilon=1$~\cite{KP17,KW20}.
As expected~\cite{MC,6loops,eps6,boot,vic-rev}, these approximate recursion relations reproduced the small values of the cubic $v^\ast_C$ and $u^\ast_C-u^\ast_I$. Since both $\lambda^I_v$ and the stability exponent of the cubic fixed point are very small, the renormalization-group flow near these fixed points is very slow. As a result, although the asymptotic values of the exponents of the two fixed points are very close, they may never be reached for reasonable experimental values of $|T-T^{}_c$.  In practice, experiments will mostly observe $t$-dependent effective exponents, e.g.,
\begin{align}
\beta^{}[u(\ell),v(\ell)]=\frac{\partial \log |\langle{\bf S}\rangle|}{\partial \log |t|},
\end{align}
valid in a temperature range around $t(0)=t(\ell)e^{-\ell/\nu}$.

\begin{figure}[htb]
\centering
\includegraphics[width=.42\textwidth]{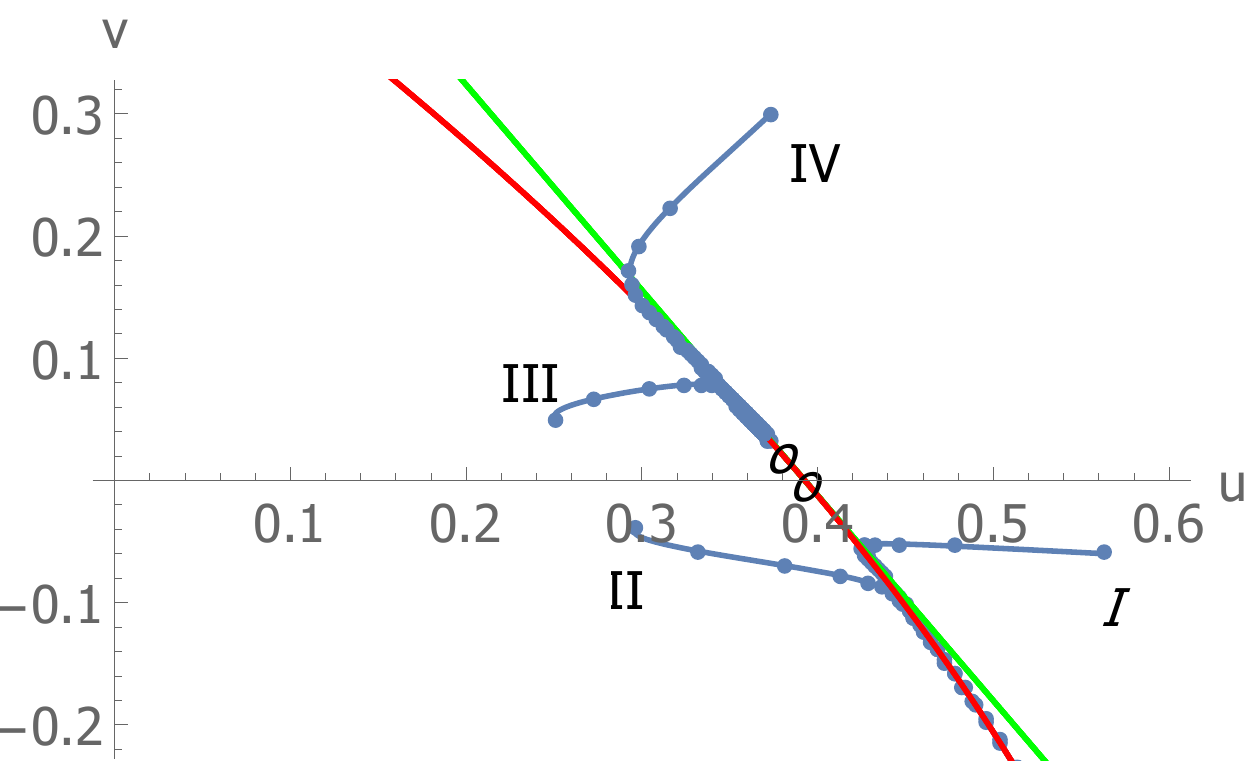}
\caption{(color online) Flow trajectories in the $u-v$ plane (blue) for several initial points of the renormalization-group iterations. The dots indicate integer values of $\ell$. The red line is the  universal asymptotic line, and the green line is the asymptotic line in the linear approximation. The small circles denote the isotropic ($v^\ast_I=0$) and cubic fixed points. After Ref. \onlinecite{AEK}.}
\label{f1}
\end{figure}

Examples of trajectories representing the renormalization-goup flows in the $u-v$ plane are reproduced in Fig. \ref{f1}. The detailed expressions for these trajectories are given in Ref. \onlinecite{AEK}.
As seen, each  trajectory has two (or three) main parts. In the first $\ell^{}_1$ iterations, the  scaling field related to $u(\ell)$ decays quickly to zero, implying a fast  non-universal transient flow  towards the universal asymptotic (red) line~\cite{AEK}. In this part, the points at integer values of $\ell$ are rather far from each other,  indicating the fast flow.  In the second part, the trajectory practically coincides with the asymptotic line. On this line, the points at integer values of $\ell$ become dense, indicating a slow variation with $\ell$. For $v>0$ (when ordering is preferred along a cubic diagonal), this implies a slow approach to the cubic fixed point. For $v<0$ (when ordering is preferred along a cubic axis), this  slow flow is followed by a third part, in which the flow gradually speeds up as the trajectory moves towards more negative values of $v$.
Eventually these trajectories are expected to cross the stability line $v=-u$, where the transition becomes first-order.
Such fluctuation-driven first-order transitions arise in many systems~\cite{1stordr,dom-muk,blank}.

Near the multicritical point the quadratic anisotropies scaling fields scale as $g(\ell)\sim g(0)e^{\lambda^{}_{axis}\ell}$ and $p(\ell)\sim p(0)e^{\lambda^{}_{diag}\ell}$, with the effective exponents $\lambda^{}_{axis}=\varphi^{}_{axis}/\nu$ and $\lambda^{}_{diag}=\varphi^{}_{diag}/\nu$ depending on $u(\ell)$ and $v(\ell)$. Except at the rotationally-invariant isotropic fixed point, these two exponents differ from each other~\cite{AAC}.
For the analysis of the multicritical phase diagrams we calculated new sixth-order expansions for these exponents in $u$ and $v$, 
using the Feynman diagrams  in Refs. \onlinecite{KP17,below}. The necessary projectors in the space of the quadratic operators were taken from  Ref.  \onlinecite{zanusso}. % In their notations, $\varphi^{}_{diag}=\theta^{}_Y/\theta^{}_S$, $\varphi^{}_{axis}=\theta^{}_X/\theta^{}_S$. 
%After that, we used expansions for a cubic fixed point from Ref. \onlinecite{eps6}.
The expansions for arbitrary
values of $n$ can be obtained, 
 as a Mathematica file, from AK.
The (universal) first and second derivatives of these series, at the isotropic fixed point were then resummed using the method described in Appendix A of Ref. \onlinecite{AEK}.
Using these resummed numbers, listed in Table \ref{I}, results in
 \begin{widetext}
\begin{eqnarray}
&\varphi^{}_{diag}(u,v)=\varphi^{I}_{}+e^{}_{10}\delta u+e^{}_{01}v+e^{}_{11}\delta u v
+[e^{}_{20}\delta u^2+e^{}_{02}v^2]/2,\nonumber\\
&\varphi^{}_{axis}(u,v)=\varphi^{I}_{}+ f^{}_{10}\delta u+f^{}_{01}v+f^{}_{11}\delta u v
+[f^{}_{20}\delta u^2+f^{}_{02}v^2]/2.
\label{effexp}
\end{eqnarray}
\end{widetext}

\begin{table}[t]
 \centering
    \caption{Numerical estimates of coefficients entering Eqs. (\ref{effexp}). The numbers are found by means of the resummation procedure described in Ref. \onlinecite{AEK}.} %
    \label{one}
     \setlength{\tabcolsep}{9.2pt}
    \begin{tabular}{clcl}
      \hline
      \hline
      Quantity & Value& Quantity & Value  \\
      \hline
    $u^\ast_I$&$0.39273(63)$& $\varphi^{I}_{}$      &$1.263(13)$\cite{KW20}         \\
    $e^{}_{10}$      &$0.928(55)$ &$e^{}_{01}$       &$0.771(40)$ \\
    $e^{}_{11}$      &$0.96(12)$ &$e^{}_{20}$       &$1.34(22)$ \\
    $e^{}_{02}$      &$0.344(25)$ & $f^{}_{10}$       &$0.928(55)$ \\
    $f^{}_{01}$      &$0.209(25)$ &$f^{}_{11}$       &$0.57(24) $\\
    $f^{}_{20}$      &$1.34(22) $ &$f^{}_{02}$       &$-0.051(87) $ \\
    \hline
    \hline
    \end{tabular}
    \label{I}
\end{table}

\begin{figure}[htb]
\includegraphics[width=0.4\textwidth]{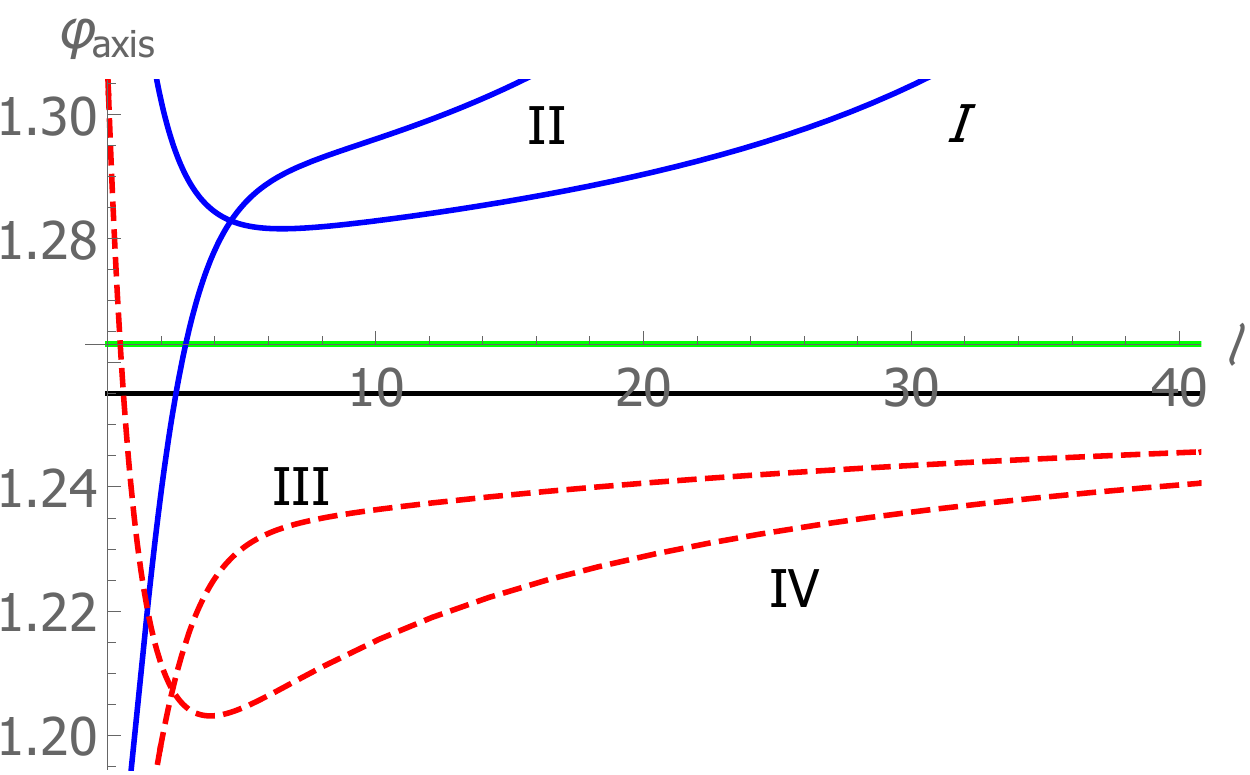}\\
\vspace{5mm}
\includegraphics[width=0.4\textwidth]{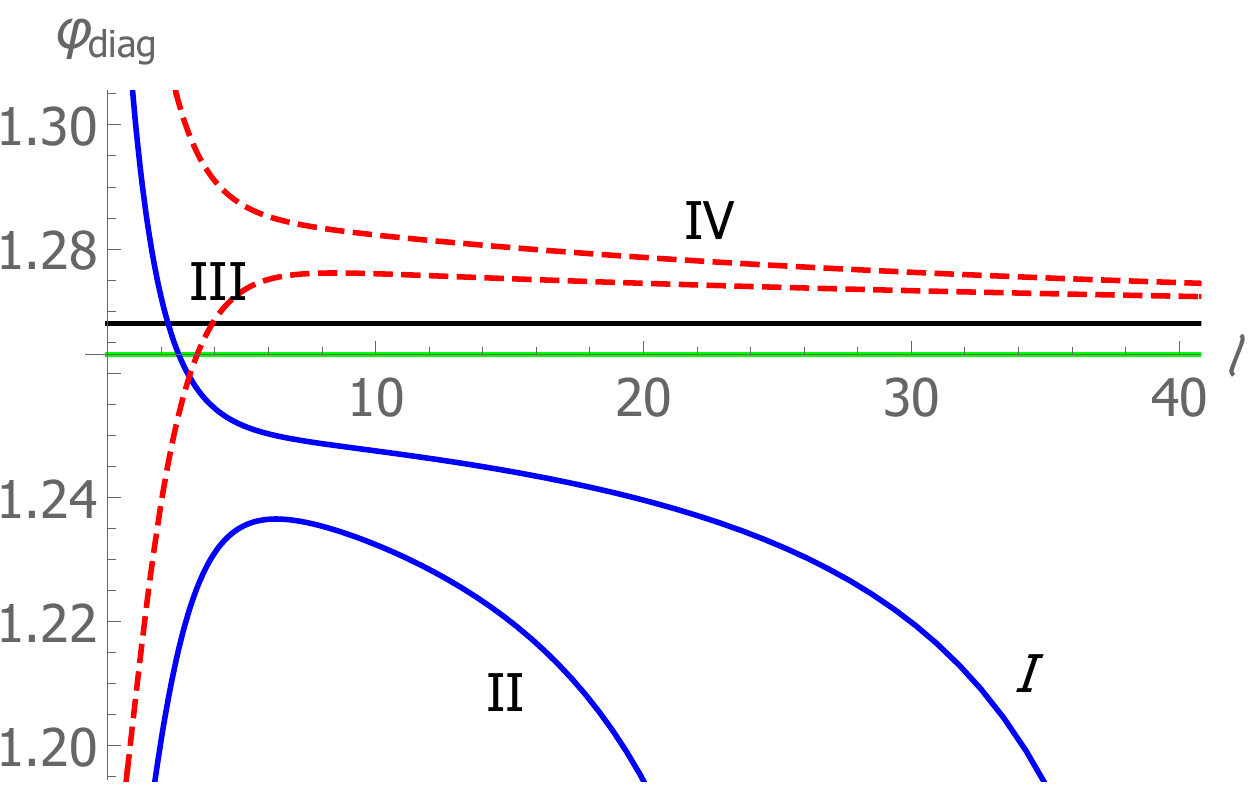}
\caption{(color online)  The effective exponents $\varphi^{}_{diag}(\ell)$ and $\varphi^{}_{axis}(\ell)$ for the trajectories shown in Fig. \ref{f1}, as functions of $\ell$. The horizontal axes (green lines) are at the asymptotic values of the isotropic fixed point, $\varphi^I=1.263$ (which is the same for both panels).  The black horizontal lines show the cubic asymptotic values, $\varphi^C_{axis}=1.255$ and $\varphi^C_{diag}=1.268$. The exponents corresponding to trajectories with $v(0)>0$ (III and IV, dashed lines) approach the asymptotic values of the cubic fixed point, visibly different from the isotropic counterparts. In contrast, those with $v(0)<0$ (I and II, full lines)  initially come close to  these values, but then  turn downward to smaller values, towards the fluctuation-driven first-order transition.  }
\label{f3}
\end{figure}

 Figures \ref{f3} depict the values of these  exponents,  calculated with $u(\ell)$ and $v(\ell)$ for the four trajectories shown in Fig. \ref{f1}. The values of the exponents at a particular value of $\ell$ are the effective values, which are expected to be observed for temperatures around $t(\ell)$. All the lines show initial (relatively fast changing) transient values, which depend on the initial values. For $v(0)>0$ (trajectories III and IV, dashed lines), the effective values then approach the asymptotic cubic exponents, from opposite directions: $\varphi^{}_{axis}$ is significantly lower than $\varphi^C_{axis}$, which in turn is below $\varphi^I$, and $\varphi^{}_{diag}$ is significantly larger than $\varphi^C_{diag}$, which in turn is above $\varphi^I$.
 For $v(0)<0$ (trajectories I and II, full lines), the results are even more significant: after the initial transient variation, the effective exponents $\varphi^{}_{axis}$ and $\varphi^{}_{diag}$ move in opposite directions away from $\varphi^I$. In each case, this variation of the effective exponents is universal: the curves of different trajectories can be collapsed onto each other by a shift in $\ell \sim \log|t|$. This results from the fact that all the trajectories flow on the same universal (red) line in Fig. \ref{f1}~\cite{AEK}.

\section{Multicritical phase diagrams for the cubic case}

Consider first Fig. \ref{1}(b), with $v>0$. The tetracritical point is described by the cubic fixed point, and at $h=0$ the only relevant fields are $t$ and $g$. The singular free energy density then obeys the scaling relation
\begin{align}
{\cal F}(t,g)=e^{-d\ell}{\cal F}\big(t(\ell),g(\ell)\big).
\end{align}
Very close to the cubic fixed point we can write
\begin{align}
t(\ell)=t e^{\ell/\nu},\ \ \ g(\ell)=g e^{\ell \lambda^C_{axis}},
\end{align}
and after $\ell^{}_f$ iterations, when $t(\ell^{}_f)\sim 1$, we end up with
\begin{align}
{\cal F}(t,g) =|t|^{d\nu}{\cal F}\big(1,g/|t|^{\varphi^C_{axis}}\big),
\label{sfe}
\end{align}
where the crossover exponent is $\varphi^C_{axis}=\nu^C\lambda^C_{axis}$.
For $g\ne 0$, the function ${\cal F}(t,g)$ must be singular on the critical line $T^{}_1(g),~g<0$, and this can happen only if the right hand side of Eq. (\ref{sfe}) has the singular form ${\cal F}(1,y)\sim (y-y^{(1)}_c)^{d\nu^I}$, with $\nu^I$ being the correlation length exponent of the Ising model~\cite{pfeuty,wegner,bruce}. Therefore this critical line occurs at
$g/|t|^{\varphi^C_{axis}}=y^{(1)}_c$, implying that
\begin{align}
T_1^{}(g)/T^{}_c(0)-1=t^{}_c(g)=(g/y^{(1)}_c)^{1/\varphi^C_{axis}}.
\label{T1g}
\end{align}
Since $\varphi^C_{axis}>1$, the shift of $T^{}_1(g)$ from $T^{}_1(0)$ approaches the $T-$axis tangentially. The same argument applies to $T'^{}_1(g)$.

As explained in Ref. \onlinecite{bruce}, the  lines $T^{}_2$ and $T'^{}_2$ exist only for $v>0$ and then they depend on $g$ via $g/v$ (except for an analytic term of order $g$).  Unlike near the isotropic fixed point, the parameter $v$ is non-zero at the cubic fixed point, $v=v^\ast_C>0$, and therefore these two critical lines also follow from singularities of ${\cal F}$ as function of $g/|t|^{\varphi^C_{axis}}$, yielding the same shift exponent, e.g.
\begin{align}
T_2^{}(g)/T^{}_c(0)-1=(g/y^{(2)}_c)^{1/\varphi^C_{axis}}.
\end{align}
However, the slow approach of $v(\ell)$ to $v^\ast_C$ may introduce slowly varying corrections. Since the asymptotic region of the cubic fixed point is reached only at very small $|t|$ (and therefore also small $|g|$), in practice the asymptotic exponents must be replaced by their effective counterparts. The measured values of $\varphi^C_{axis}$ are therefore expected to be smaller than $\varphi^C_{axis}<\varphi^I$ [Fig. \ref{f3}].

We now turn to the bicritical phase diagram, Fig. \ref{1}(a). For $v>0$, this diagram applies to the diagonal uniaxial anisotropy (\ref{Hdiag}), with the order-parameters along $[111]$ ($p>0$) or in mixed phase ($p<0$). The two critical lines still have the form (\ref{T1g}), but now with the effective exponent $\varphi^{}_{diag}$, which is expected to be larger than the asymptotic $\varphi^C_{axis}>\varphi^I$.
The  bicritical phase diagram, Fig. \ref{1}(a), changes considerably for for $v<0$ and ordering along an axis. In this case, the bicritical point is related to the renormalization-group flow towards the fluctuation-driven first-order transition. If the measurement is performed at `intermediate' values of $|t|$, before the first-order borderline is crossed, then the phase diagram may still look the same as Fig. \ref{1}(a), but with effective exponents $\varphi^{}_{axis}$ which gradually increase significantly as $|t|$ decreases [Fig. \ref{f3}].    
For smaller $|t|$, the  bicritical point will be replaced by a triple point: the transitions at $T^{}_1(g)$ and $T'^{}_1(g)$ will become first-order in its vicinity, and the three first-order transition lines meet at that point. For larger $|g|$ these transitions become second order, beyond tricritical points~\cite{dom-muk}. The first and second order transitions on these lines become even more complicated when one adds the dangerously irrelevant terms of order $|{\bf S}|^6$~\cite{blank}, or when one mixes the two quadratic anisotropies (\ref{Hg}) and (\ref{Hdiag})~\cite{ker}.

\section{The biconical (non-cubic) case}

Both Eqs. (\ref{Hv}) and (\ref{n1n2}) are special cases of the
general Hamiltonian, with only even powers of the order-parameter components,
\begin{align}
{\cal H}=\int d^dr\big[{\boldmath{\nabla}}{\bf S}({\bf r})|^2/2+U[{\bf S}({\bf r})\big],
\end{align}
with
\begin{align}
U[{\bf S}]=\sum_{i,j=1}^n r^{}_{ij}S^{}_iS^{}_j/2+\sum_{i,j,k,l=1}^n u^{}_{ijkl}S^{}_iS^{}_jS^{}_kS^{}_l.
\end{align}
The isotropic Hamiltonian is given by $r^{}_{ij}=r\delta^{}_{ij}$ and $u^{}_{ijkl}=u[\delta^{}_{ij}\delta^{}_{kl}+\delta^{}_{ik}\delta^{}_{jl}+\delta^{}_{il}\delta^{}_{jk}]/3$.
This yields the isotropic (Heisenberg) fixed point.

The  terms in $U$ involve combinations of the order-parameter components products, which form irreducible representations of the ${\cal O}(n)$ internal group~\cite{gr,brezin,michel}. At order $m$ in the spin components, these can be written as~\cite{MC,vicari,wegner}
\begin{align}
P^{a^{}_1,\dots,a^{}_\ell}_{m,\ell}=|{\bf S}|^{m-\ell}Q_\ell^{a^{}_1,\dots,a^{}_\ell}({\bf S}),
\end{align}
where $Q_\ell^{a^{}_1,...,a^{}_\ell}({\bf S})$ is a homogeneous polynomial of degree $\ell$, symmetric and traceless in the $\ell$ indices.  Under the renormalization-group flow the operators with different
$\ell$  never mix, and all the terms associated with the same values of $m,\ell$  have the same renormalization-group  stability exponent near the isotropic fixed point, $\lambda^{}_{m,\ell}$. The exponent $\lambda^{}_{m,0}$ corresponds to the isotropic term $|{\bf S}|^m$, and all the other exponents are associated with traceless symmetry-breaking terms.

For the quadratic terms, the only symmetry-breaking terms are the traceless combinations
\begin{align}
P^{ij}_{2,2}=Q^{ij}_{2}=g^{ij}[S^{}_iS^{}_j-\delta^{}_{ij}{\bf S}^2/n].
\end{align}
Equations (\ref{Hg}) and (\ref{Hdiag}) are special cases of this expression, for axial and diagonal symmetry-breaking.
 All these terms have the same crossover exponent $\varphi^I=\lambda^{I}_{2,2}\nu$ at the isotropic fixed point, but the axial and diagonal crossover exponents  $\varphi^{}_{axis}$ and $\varphi^{}_{diag}$ differ from each other for non-isotropic fixed points~\cite{AAC}.

The quartic traceless terms involve only two families of $P$'s, $P^{ijkl}_{4,4}$ and $P^{ij}_{4,2}$, and all the traceless  anisotropic terms are linear combinations of them. %, with the exponents (near the isotropic fixed point) $\lambda^{}_{4,4}$ and $lambda^{}_{4,2}$.
The cubic potential has the form~\cite{MC,vicari}
\begin{align}
\sum_{i=1}^n(S_i)^4=\sum_{i=1}^nP_{4,4}^{iiii}\big({\bf S})+\frac{3}{n+2}P^{}_{4,0}({\bf S}\big),
\end{align}
and thus $\lambda^I_v=\lambda^{}_{4,4}$.

The model including  Eqs. (\ref{Hg}) and (\ref{n1n2}) involves the quadratic term
\begin{align}
O^{}_{2,2}=[{\bf S}^{}_1]^2-\frac{n^{}_1}{n}{\bf S}^2=\sum_{i=1}^{n^{}_1} P_{2,2}^{i,i}
\end{align}
and the  quartic terms (\ref{n1n2}), which can be written as~\cite{MC,vicari}
\begin{align}
{\cal H}^{}_{n^{}_1-n^{}_2}\equiv g^{}_0 O^{}_{4,0}+g^{}_2O^{}_{4,2}+g^{}_4O^{}_{4,4},
\end{align}
where $O^{}_{4,0}\equiv P^{}_{4,0}=|{\bf S}|^4$ is the isotropic term, and
\begin{align}
&O^{}_{4,4}=\sum_{i=1}^{n^{}_1}\sum_{j=n^{}_1+1}^{n}P_{4,4}^{iijj}=[{\bf S}^{}_1]^2[{\bf S}^{}_2]^2\nonumber\\
&-\frac{1}{n+4}{\bf S}^2(n^{}_1[{\bf S}^{}_2]^2+n^{}_2[{\bf S}^{}_1]^2)+\frac{n^{}_1n^{}_2}{(n+2)(n+4)}[{\bf S}^2]^2,\nonumber\\
&O^{}_{4,2}={\bf S}^2O^{}_{2,2}.
\end{align}
Near the isotropic fixed point these traceless terms have  the exponents $\lambda^{}_{2,2}\equiv \lambda^I_g$, $\lambda^{}_{4,4}\equiv\lambda^I_v$ and $\lambda^{}_{4,2}$. The latter two exponents determine the scaling of any quartic traceless perturbation.
 Field-theory  calculations give $\lambda^{}_{4,2}\approx -0.55$~\cite{vicari}, and therefore $g^{}_2$ is irrelevant near the isotropic fixed point (and probably also near the cubic and the biconical fixed points, which are very close). Since $\lambda^I_v\gtrapprox 0$ at $n=d=3$, the
  isotropic fixed point is unstable for both models. For the cubic model $g^{}_2=0$, and the renormalization-group trajectories are all in the $u-v$ plane, with the stable cubic fixed point very close to the isotropic one, as discussed in Sec. II.

 For the (\ref{n1n2}) model we also need $g^{}_2$, and the rennormalization-group trajectories are in the three-dimensional space of $u^{}_1,~u^{}_2$ and $w$ (or $g^{}_0,~g^{}_2$ and $g^{}_4$). In this case one ends up with six fixed points: Gaussian ($u^\ast_1=u^\ast_2=w^\ast=0)$, $n^{}_1~(u^{\ast}_1>0~u^{\ast}_2=w^\ast=0)$, $n^{}_2~(u^{\ast}_2>0,~u^{\ast}_1=w^\ast=0)$, Decoupled ($u^\ast_1,~u^\ast_2>0,~w^\ast=0)$, Isotropic ($u^\ast=u^\ast_2=w^\ast>0)$ and Biconical (all three coefficients different and non-zero)~\cite{13,MC,vicari}. At $d=n=1+2=3$, only the biconical fixed point is stable. Since the stability exponent of the isotropic fixed point is very small, the biconical fixed point is also expected to be very close to the isotropic one. Indeed, this is the result of the 
 field-theoretical calculations,  with the quartic stability exponents $\lambda^{}_{3,4,5}\approx -0.583,~-0.554,~-0.01$~\cite{folk}, and with asymptotic exponents which are very close to those of the isotropic fixed point. Note however that the anisotropic biconical fixed point has different exponents for quantities associated with $S^{}_1\equiv S^{}_\parallel$ (e.g., $\beta^B_\parallel$ and $\gamma^B_{\parallel}$) and with ${\bf S}^{}_2\equiv {\bf S}^{}_\perp$ (e.g., $\beta^B_\perp$ and $\gamma^B_{\perp}$). Although these have asymptotic biconical values close to those of the isotropic fixed point, their effective values away from criticality may be quite different (similarly to Fig. \ref{f3}). The asymptotic biconical crossover exponent $\varphi^B$ was also found to be close to $\varphi^I$, but its effective values may differ. Reference \onlinecite{folk} calculated flow diagrams and effective exponents for this case, but used only second order in $\epsilon$. It will be interesting to repeat our analysis of Ref. \onlinecite{AEK} also for the competition between the isotropic and the biconical fixed points. Generally,
   we expect renormalization-group flows similar to those described in Sec. II: after some fast transient flows, in which the non-linear scaling fields associated with the irrelevant parameters $\delta g^{}_0=g^{}_0-g^\ast_I$ and $g^{}_2$ decay to zero, the flow will reach a universal line (in the three-dimensional critical space), along which it will slowly approach the biconical fixed point for $w^2<u^{}_1u^{}_2$, with a tetracritical phase diagram,  or flow towards a fluctuation-driven first-order transition (for $w^2>u^{}_1u^{}_2$), with an intermediate bicritical phase diagram (at intermediate values of $t$ and $g$).

 As far as we know, there  does not yet exist an accurate analysis which combines the Hamiltonians (\ref{Hv}) and (\ref{n1n2}), requiring the four dimensional space of $\delta g^{}_0,~g_2,~g^{}_4$ and $v$. The last two parameters are slightly relevant near the isotropic fixed point, with the same exponent $\lambda^I_v$. However, the above discussion indicates that although this combined model generates generalized flows, the qualitative implications on the multicritical phase diagrams may not change.

\section{Experiments}

The phase diagrams in Fig. \ref{1} have been observed for the structural phase transitions in the perovskites, as reviewed in Ref. \onlinecite{AEK}. For SrTiO$^{}_3$ ($v<0$), stressed along $[100]$ (Fig. \ref{1}(a)], Stokka and Fossheim~\cite{stokka} found a `regular' bicritical phase diagram, as in Fig. \ref{1}(a), without the triple poit expected for very large $\ell$.  This is probably due to the very small value of $v(0)$. They measured $\varphi^{}_{axis}=1.27\pm .06$, above $\varphi^I_{axis}$.
For LaAlO$^{}_3$ ($v>0$), stressed along $[111]$ (Fig. \ref{1}(a) but with ordering along $[111]$ for $g<0$ and with the mixed phase for $g>0$), M\"{u}ller {\it et al.}~\cite{KAM-NATO} found
$\varphi^{}_{diag}=1.31\pm.07$, higher than the asymptotic $\varphi^C_{diag}$. Both results are  in qualitative agreement with the effective exponents in Fig. \ref{f3}.

The $T-g$ diagrams in Fig. \ref{1} were also observed in the uniaxial antiferromagnet with a longitudinal and a transverse magnetic fields, with $n=1+2$~\cite{king,Shapira}. Similar bicritical phase diagrams were also found in Monte Carlo simulations of the XXZ model~\cite{selke,landau}. Surprisingly, these experiments gave a bicritical phase diagrams, although the stable biconical fixed point would imply a tetracritical point. Alternatively, if the initial Hamiltonian was out of the region of attraction of the biconical fixed point then the bicritical point should have turned into a triple point, with two tricritical points on the lines $T^{}_1(g)$ and $t'^()_1(g)$ in Fig. \ref{1}(a)~\cite{dom-muk}. Such points were not seen in the experiments and in the simulations. This can be explained by our scenario:  for the relatively large temperature range or the relatively small finite sizes, the slow renormalization-group flows stay in the vicinity of the isotropic fixed point, with the effective behavior of this vicinity.
Indeed, for tetragonal MnF$^{}_2$, King and Rohrer~\cite{king} measured $\varphi^{}_{axis}=1.279\pm.031$. For cubic RbMnF$^{}_3$, with $v>0$, and a magnetic field along $[100]$, one expects Fig. \ref{1}(b). Shapira~\cite{Shapira} cites $\varphi^{}_{axis}=1.258\pm.08$ and $\varphi^{}_{axis}=1.278\pm.026$ or $1.274\pm.045$. These values are somewhat larger than what we find in Fig. \ref{f3}, but the effects of the cubic anisotropy on the magnetic order-parameters may be small, and the critical behavior may be dominated by the effective exponents related to the three-dimensional flows discussed in Sec. IV.

The exponents $\varphi^{}_{axis}$ and $\varphi^{}_{diag}$ can also be measured at $g=0$ or $p=0$. %,   by measuring the derivative
For the structural transitions in the perovskites~\cite{10,luthi}, the  order-parameter components couple to the strain degrees of freedom via
\begin{align}
{\cal H}^{}_{el}=-B\sum_{i=1}^3e^{}_{ii}[(S^{}_i)^2-{\bf S}^2/3]%\nonumber\\%-B^{}_2\{e^{}_1[(S^{}_2)^2+(S^{}_3)^2]\nonumber\\
%+e^{}_2[(S^{}_1)^2+(S^{}_3)^2]+e^{}_3[(S^{}_1)^2+(S^{}_2)^2]\}\nonumber\\
-B'\sum_{i<j}e^{}_{ij}S^{}_iS^{}_j,
\label{Hel}
\end{align}
where $\{e^{}_{ij}\}$ are the strains.
In linear response, the strains can be replaced by $e^{}_{i,j}=c^{}_{ij} \sigma^{}_{ij}$, with the corresponding uniaxial stresses $\sigma^{}_{ij}$, and with shifts in $u$ and $v$. In our Eqs. (\ref{Hg}) and (\ref{Hdiag}), $g= \sigma^{}_{11}$ and $-p=\sigma^{}_{ij},~i\ne j$.
Therefore, the jump in the strains at $g=0$ is found from Eq. (\ref{sfe}) to be~\cite{10}
\begin{align}
\langle e^{}_{11}\rangle=\frac{\partial {\cal F}}{\partial \sigma^{}_{11}}\sim \langle (S^{}_1)^2-{\bf S}^2/3\rangle\sim |t|^{\tilde{\beta}^{}_{axis}},
\end{align}
with $\tilde{\beta}^{}_{axis}=d\nu-\varphi^{}_{axis}$. The elastic compliance is the second derivative, which diverges as $|t|^{-\tilde{\gamma}^{}_{axis}}$, with $\tilde{\gamma}^{}_{axis}=2\varphi^{}_{axis}-d\nu$.
Similarly, 
\begin{align}
\langle e^{}_{12}\rangle=\frac{\partial {\cal F}}{\partial \sigma^{}_{12}}\sim \langle S^{}_1S^{}_2\rangle\sim |t|^{\tilde{\beta}^{}_{diag}},
\end{align}
with $\tilde{\beta}^{}_{diag}=d\nu-\varphi^{}_{diag}$, and with the corresponding compliance exponent  $\tilde{\gamma}^{}_{diag}=2\varphi^{}_{diag}-d\nu$.

Experiments give a wide range of results~\cite{hochli}.  Rehwald~\cite{rehwald} measured the two compliances for SrTiO$^{}_3$, which has a small $v<0$, and found $\tilde{\gamma}^{}_{axis}\approx 0.69,~ \tilde{\gamma}^{}_{diag}\approx 0.62$, equivalent to $\varphi^{}_{axis}\approx 1.31,~\varphi^{}_{diag}\approx 1.29$. Replacing $d\nu$ by $2-\alpha$, L\"{u}thi and Rehwald~\cite{luthi,KAMrev} estimated $\varphi^{}_{axis}\approx 1.4$, in agreement with their measurement of the shift exponent near the bicritical point. These values for the axial case agree with the (large) effective values in Fig. \ref{f3}, but the diagonal exponent seems too high. Under uniaxial stress both exponents decreased, apparently indicating the crossover from $n=3$ to $n=1$ or $n=2$. For ferroelectric BaTiO$^{}_3$ they cite smaller values, but these must correspond to the cubic dipolar case, for which we do not yet have accurate predictions~\cite{AAdip}. Ultrasonic experiments on KMnF$^{}_3$~\cite{holt}, before its first-order transition,  gave $\varphi^{}_{axis}\approx 1.26$. Since $v$ has a more negative value here, this effective value  probably corresponds to the small$-\ell$ values in Fig. \ref{f3}.

For the uniaxial antiferromagnets with a longitudinal magnetic field, the above strains are replaced by $\sigma^{}_{ij}\rightarrow H^{}_iH^{}_j$.  The discontinuity in the magnetization across the spin-flop first-order transition line and the magnetic susceptibility near the bicritical point  scale as~\cite{13}
\begin{align}
\Delta M^{}_\parallel\sim |t|^{\tilde{\beta}^{}_{axis}},\ \ \ \chi\sim |t|^{-\tilde{\gamma}^{}_{axis}},
\end{align}
with the same exponents as above. Experimentally, the bicritical exponents deduced from these measurements on MnF$^{}_2$ and RbMnF$^{}_3$, agree with those deduced from the bicritical phase diagrams, quoted above.

In addition to critical exponents, there also exist many universal amplitude ratios~\cite{amp}. Examples include the ratio $w^{}_\parallel/w^{}_\perp$, where $\tilde{g}^{}_c=w^{}_\parallel \tilde{t}^{}_1(g)^{\varphi}$ and $\tilde{g}^{}_c=w^{}_\perp \tilde{t}'^{}_1(g)^{\varphi}$, with the appropriate scaling fields $\tilde{t}=t+q g,~ \tilde(g)=g-p t$~\cite{MEFbi}. The effective values which should replace such universal values for intermediate values of $|t|$ have not been calculated yet.

\section{Conclusions}

Our main conclusion is that the slow (universal) renormalization group flows in the vicinity of the isotropic fixed point generate effective behaviors that yield bicritical points with effective critical exponents, rather than the asymptotic tetracritical phase diagram or the asymptotic phase diagram with a triple point. This is qualitatively supported by experiments and simulations. However, dedicated experiments on varying temperature ranges may confirm our detailed estimates of the effective critical exponents.

Our recursion relations for the cubic problem near the isotropic fixed point can be extended to the more general quartic Hamiltonians, thus giving predictions for the effective multicritical phase diagrams also for the more general cases. This requires reliable high-order expansions of the derivatives of the beta-functions in the three- or four-dimensional parameter spaces, appropriate resummations of these series at $\epsilon=1$, and then solutions of the recursion relations, followed by calculations of the various $\ell-$dependent effective exponents. 

In addition to the effective exponents, it will be also useful to calculate effective values of amplitude ratios, which may deviate from their asymptotic universl values.

\begin{acknowledgments}
This paper is dedicated to the memory of Mark Ya. Azbel, who was our colleague at Tel Aviv University for many years, on his 90th birthday.
This research was initiated by stimulating discussions with Slava Rychkov.  A.K. gratefully acknowledges Mikhail Kompaniets for the helpful discussion and Andrey Pikelner for his help with RG expansions from Ref.~\onlinecite{below}, and
 the support of Foundation for the Advancement of Theoretical Physics "BASIS" through Grant 18-1-2-43-1.

\end{acknowledgments}


\begin{thebibliography}{99}
\bibitem{fn} The name `biconical' should not be confused with the name `bicritical'. In fct, the biconical fixed point represent a tetracritical point in the $T-g$ phase diagram.
\end{thebibliography}
\end{document}